\documentclass[aps,prl,twocolumn,showpacs]{revtex4}

\usepackage{amsmath,amsfonts,epsfig,graphicx,float}

\begin{document}

\title{Interaction of a Contact Resonance of Microspheres with Surface Acoustic Waves}

\author{N. Boechler$^{1}$, J. K. Eliason$^{2}$, A. Kumar$^{1}$, A. A. Maznev$^{2}$, K. A. Nelson$^{2}$, and N. Fang$^{1}$}

\affiliation{ 
$^1$ Department of Mechanical Engineering, Massachusetts Institute of Technology, Cambridge, MA 02139 \\
$^2$ Department of Chemistry, Massachusetts Institute of Technology, Cambridge, MA 02139 
}


\begin{abstract}
We study the interaction of surface acoustic waves (SAWs) with the contact-based, axial vibrational resonance of $1$~$\mu$m silica microspheres forming a two-dimensional granular crystal adhered to a substrate. The laser-induced transient grating technique is used to excite SAWs and measure their dispersion. The measured dispersion curves exhibit ``avoided crossing'' behavior due to the hybridization of the SAWs with the microsphere resonance. We compare the measured dispersion curves with those predicted by our analytical model, and find excellent agreement. The approach presented can be used to study the contact mechanics and adhesion of micro- and nanoparticles as well as the dynamics of microscale granular crystals.  

\end{abstract}




\maketitle

Wave phenomena in granular media is a rich and rapidly developing field of research~\cite{NesterenkoBook,GranularPhysicsBook,GranularCrystalReviewChapter}. At the heart of this field is the Hertzian model of elastic contact between spherical particles, in which the stiffness of the contact depends on the applied force \cite{Hertz}. One type of granular media, often referred to as ``granular crystals'', consists of close-packed, ordered arrays of elastic particles that interact via Hertzian contact \cite{NesterenkoBook,GranularCrystalReviewChapter}. Granular crystals have been shown to support a wide range of linear and nonlinear dynamical phenomena not encountered in conventional materials, and have been suggested for various engineering applications \cite{NesterenkoBook,GranularCrystalReviewChapter}.  

Acoustic studies of granular media typically involve macroscopic particles with dimensions of $\sim0.1$--$10$~mm~\cite{NesterenkoBook,GranularPhysicsBook,GranularCrystalReviewChapter}, whereas contact-based vibrations of microparticles with dimensions of (or under) $\sim1$ $\mu$m remain largely unexplored. The scale factor is significant as a microparticle system cannot be thought of simply as a scaled down version of a macroscale system which is governed by the same physics. Rather, microparticles are expected to yield qualitatively different dynamics. One crucial factor is the role of adhesion \cite{Bhushan,Israelachvili}, which is almost negligible on millimeter scales but significant on micron scales. Because of adhesion, a microsphere in contact with a substrate is pulled toward the latter. This results in an equilibrium contact stiffness and an axial ``contact resonance'' vibrational mode with frequency determined by the particle mass, the adhesion, and the elastic properties of the particle and substrate \cite{Peri2005}. This phenomenon has not hitherto been observed experimentally, although rocking-mode vibrations at much lower frequencies have been studied in $\sim5$--$50$ $\mu$m spheres \cite{Dybwad1985,Peri2005,Butt2010}, and vibrations corresponding to free particle eigenmodes at much higher frequencies have been studied in $5$ $\mu$m silica spheres~\cite{Tournat2009}.   

In this letter, we study the contact-based, axial vibrational resonance of $1$ $\mu$m diameter microspheres forming a two-dimensional granular crystal adhered to a substrate using another hitherto unexplored  phenomenon, i.e., the interaction of axial contact resonances of microparticles with surface acoustic waves (SAWs) in the substrate. We use the laser-induced transient grating (TG) technique \cite{RogersReview2000,NelsonTG2012} to excite the long-wavelength (relative to the particle size) SAWs and measure their dispersion. The measured dispersion curves exhibit classic ``avoided crossing'' behavior due to the hybridization of the SAWs with the contact resonance of microspheres. Such coupling between SAWs and mechanical surface oscillators was studied in theoretical works \cite{Kosevich1989,Baghai1992,Garova1999}, and the experimental work \cite{Laude2011}.  We analyze our measurements using a simple model that yields an analytical expression for the dispersion relation. We find excellent agreement between our model and the measurements using the frequency of the contact resonance as a single fitting parameter, and we compare our results with estimates based on the Derjaguin-Muller-Toporov (DMT) contact model \cite{DMT1983}. 
\begin{figure}[H]
\begin{center}
\includegraphics[width=8.6cm,height=6.45cm]{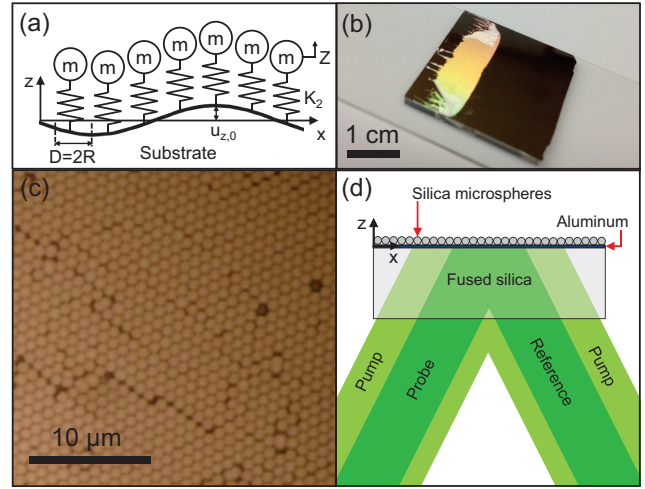}
\end{center}
\caption{\label{Figure1}[Color online] (a) Microspheres interacting with a SAW via contact ``springs'', with notations used in the theoretical model. (b) Photograph of the sample. (c) Representative image of the silica microsphere monolayer. (d) Schematic illustration of the TG setup.}
\end{figure} 
%


Our sample is a hexagonally packed monolayer of $D=1.08$ $\mu$m diameter silica microspheres deposited on an aluminum-coated fused silica substrate, as shown in Fig.~\ref{Figure1}(b,c). The fused silica slab is $1.5$~mm thick, and the aluminum layer, which serves as a medium to absorb pump laser light, is $0.20$~$\mu$m thick. To assemble the monolayer on the substrate, we used the ``wedge-shaped cell'' convective self-assembly technique \cite{SunWedgeMethod2010,LopezWedgeMethod2011,Supplementary}. The resulting monolayer has an overall area of $\sim5$~mm by $10$~mm, as shown in Fig.~\ref{Figure1}(b). A representative optical image of the monolayer packing is shown in Fig.~\ref{Figure1}(c). Although the monolayer has a relatively uniform distribution, defects and grains of uniform packing are clearly present. 


We use a laser-induced TG technique \cite{RogersReview2000,NelsonTG2012} to measure the phase velocity dispersion of SAWs in our sample. The TG setup with heterodyne detection used for these experiments has been described previously \cite{NelsonTG2012}. In summary, two excitation beams derived from the same laser source ($515$~nm wavelength, $60$~ps pulse duration, $2.44$~$\mu$J total pulse energy at the sample) enter the sample through the transparent silica substrate, as shown in Fig.~\ref{Figure1}(d), and are overlapped at the aluminum layer forming a spatially periodic interference pattern. The pump spot has $500$~$\mu$m diameter at $1/e^2$ intensity level. Absorption of the laser light by the aluminum film induces rapid thermal expansion, which leads to  the generation of counter-propagating SAWs with wavelength $\lambda_S$ defined by the period of the interference pattern. The period is controlled by a phase mask pattern used to create the two excitation beams, by splitting the incident beam into $+/-1$ diffraction orders \cite{RogersReview2000}. Switching phase mask patterns allows measurements at multiple acoustic wavelengths. 

SAW detection is accomplished using a quasi-cw probe beam ($532$~nm wavelength, $10.7$~mW average power at the sample) focused at the center of the excitation pattern to a spot of $150$~$\mu$m diameter. The probe beam also enters the sample through the silica substrate and is diffracted by surface ripples and refractive index variations in the substrate induced by SAWs~\cite{NoteSurfaceRipples}. The diffracted beam is overlapped with the reflected reference beam (local oscillator) \cite{RogersReview2000,NelsonTG2012} and directed onto a fast avalanche photodiode with a 1 GHz bandwidth. The signal is recorded using an oscilloscope and averaged over $10^4$ repetitions.  

\begin{figure}[h]
\begin{center}
\includegraphics[width=8.6cm,height=6.45cm]{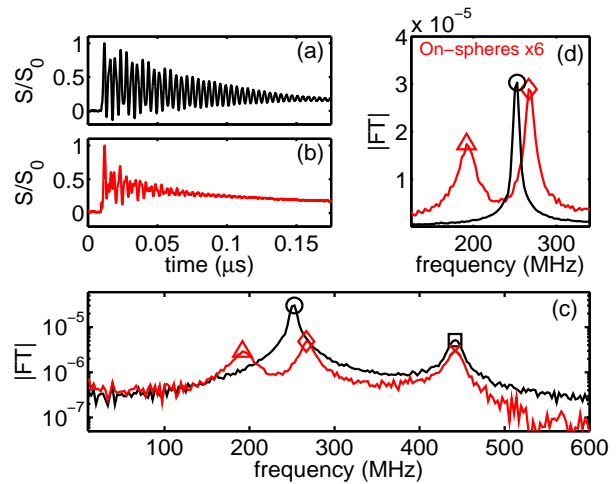}
\end{center}
\caption{\label{Figure2}[Color online] Normalized signal (a) off-spheres, and (b) on-spheres, for $k=0.46$~$\mu$$\text{m}^{-1}$. The acquired signal $S$ is normalized by the maximum signal amplitude $S_0$. Fourier transform (FT) magnitudes corresponding to the signals in (a) and (b), plotted in log scale (c) and linear scale (d). The black curve corresponds to the signal in (a), and the red curve to the signal in (b). The markers denote the identified peaks, which are plotted in Fig.~\ref{Figure3} using the same markers.}
\end{figure}

Figure~\ref{Figure2}(a,b) shows typical signal waveforms acquired at an acoustic wave vector magnitude $k=2\pi/ \lambda_S=0.46$~$\mu$$\text{m}^{-1}$. Figure~\ref{Figure2}(a) shows the signal from a sample location without microspheres, and Fig.~\ref{Figure2}(b) corresponds to a location with spheres. In both cases there is a sharp initial increase, which corresponds to the excitation pulse arriving at the sample. The slowly decaying component is due to the ``thermal grating'' associated with the temperature profile in the sample \cite{RogersReview2000,NelsonTG2012}. The high frequency oscillations are due to acoustic waves.

\begin{figure}[h]
\begin{center}
\includegraphics[width=8.6cm,height=6.45cm]{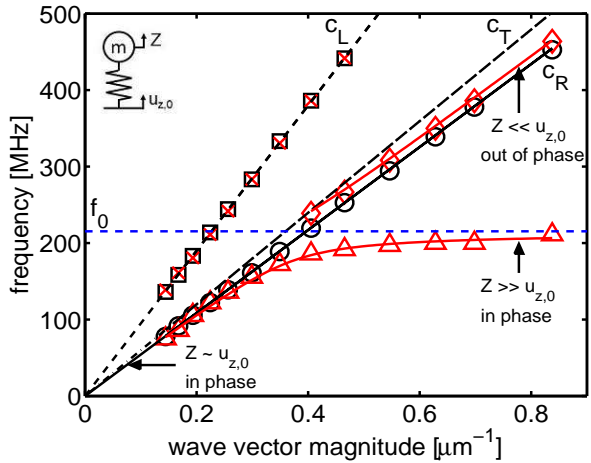}
\end{center}
\caption{\label{Figure3}[Color online] Dispersion relations. Red and black markers are the measured frequency peaks for the on- and off-spheres cases, respectivly. The solid red line is the dispersion calculated using our model. Also shown are lines corresponding to longitudinal, transverse, and Rayleigh waves in fused silica, and a horizontal line corresponding to the microsphere contact resonance frequency.}
\end{figure}

Figure~\ref{Figure2}(c,d) shows the Fourier spectra of acoustic oscillations corresponding to the signals in Fig.~\ref{Figure2}(a,b) \cite{Supplementary}. In the off-spheres case there are two clear peaks, which correspond to a Rayleigh SAW (the low frequency peak) and a longitudinal wave in the substrate~\cite{NoteLateralWaves}. Figure~\ref{Figure3} shows the acoustic dispersion curves. For the off-spheres case, we see linear dispersion curves that agree very well with lines corresponding to the longitudinal and Rayleigh wave velocities in fused silica. We used typical wave speeds for fused silica of $c_L=5968$~m/s (longitudinal) and $c_T=3764$ m/s (transverse)~\cite{Silica}, and we calculated the Rayleigh wave speed $c_R=3409$ m/s by numerically solving the Rayleigh equation~\cite{Ewing}. More accurate calculations accounting for the aluminum layer~\cite{Supplementary,Ewing} showed that the reduction in the Rayleigh wave speed due to the aluminum layer does not exceed $1.4\%$. 

The on-spheres case yields starkly different behavior from the off-spheres case, as can be seen by comparing the signal waveforms shown in Fig. 2(a,b). The comparison of the spectra in Fig. 2(c) shows that the longitudinal peak is unaffected by the presence of the spheres whereas the SAW peak is split in two. The on-spheres dispersion curves in Fig.~\ref{Figure3} reveal a classic ``avoided crossing'' between the Rayleigh wave and a local microsphere resonance.  The lower branch starts as a Rayleigh wave at low wave vector magnitudes and approaches a horizontal asymptote corresponding to the resonance frequency. The upper branch is close to the Rayleigh wave at high wave vector magnitudes; in the ``avoided crossing'' region it deviates from the Rayleigh line and stops at the threshold corresponding to the transverse acoustic velocity of the substrate.  
%

To describe SAW propagation in our system of microspheres coupled to an elastic substrate, we adapt an approach developed in the theoretical works \cite{Kosevich1989,Baghai1992,Garova1999}. As shown schematically in Fig.~\ref{Figure1}(a), we model the substrate as an elastic half-space ($z\le0$), where the surface of the halfspace is connected to an array of linear surface oscillators, with mass $m$ and stiffness $K_2$, which represent the microspheres connected to the substrate via ``contact springs''. We calculate the mass of a microsphere as $m=\frac{4}{3}\pi(\frac{D}{2})^3 \rho_s$, where $\rho_s=2.0$~g/$\text{cm}^3$ is the density of the silica microspheres as provided by the manufacturer (Corpuscular, Inc.).

We approximate the microspheres as point-masses, as the lowest spheroidal resonance of the microspheres, $f_{2}=2.9$ GHz \cite{Sato}, is much greater than the frequencies observed in the experiment. The equation of motion for the surface oscillator can be written as $m\ddot{Z}+K_2(Z-u_{z,0})=0$, where $u_{z,0}$ is the displacement of the substrate surface, and $Z$ is the displacement of the oscillator relative to the surface. The particles exert a vertical force on the substrate, leading to the following boundary conditions at the surface $z=0$:
\begin{equation}
\begin{split}
\sigma_{zz}=\frac{K_2(Z-u_{z,0})}{A} 
\qquad
\sigma_{xz}=0, 
\end{split}
\label{BCs}
\end{equation}
where $\sigma_{zz}$ and $\sigma_{xz}$ are components of the elastic stress tensor \cite{Ewing}, and $A=\frac{\sqrt{3}D^2}{2}$ is the area of a primitive unit cell in our hexagonally packed monolayer. 

Since the acoustic wavelength is much larger than the sphere size, in Eq.~\ref{BCs} we use an effective medium approach, and approximate the average normal stress at the surface as the force exerted by the microsphere ``spring'' divided by the area of a unit cell (assuming perfect packing) even though the stress is localized around the contact area. We follow the standard procedure for the derivation of the Rayleigh wave equation \cite{Ewing}, but use Eq. \ref{BCs} instead of stress-free boundary conditions to obtain the following dispersion relation for the SAWs in our coupled oscillator system \cite{Supplementary}: 
\begin{widetext} 
\begin{equation}	(\frac{\omega^2}{\omega_0^2}-1)[(2-\frac{{\omega}^2}{{k^2 c_T}^2})^2-4(1-\frac{{\omega}^2}{{k^2 c_L}^2})^{1/2}(1-\frac{{\omega}^2}{{k^2 c_T}^2})^{1/2}]=\frac{m}{A\rho_2}\frac{\omega^4(1-\frac{\omega^2}{k^2 c_L^2})^{1/2}}{k^3 c_T^4},
\label{RayleighOscEqn}
\end{equation}    
\end{widetext}
where $\omega_0=2\pi f_0=\sqrt{K_2/m}$ is the angular frequency of the contact resonance, and $\rho_2=2.2$~g/$\text{cm}^3$~\cite{ASMSilica} is the density of the silica substrate. On the left-hand side of Eq.~\ref{RayleighOscEqn}, the term in square brackets is familiar from the Rayleigh equation, and the term in the parentheses describes the resonance of the oscillators. The right-hand side of Eq.~\ref{RayleighOscEqn} is responsible for the coupling between the Rayleigh waves and the oscillators; if it is made to vanish (for instance by assuming a vanishing areal density $m/A$), then the oscillators and SAWs in the substrate are effectively decoupled. We also see that $A$ and $\rho_2$ relate to the coupling strength, $K_2$ relates to the frequency of the avoided crossing through $\omega_0$, and $m$ relates to both.


By taking the frequency of the contact resonance as a fitting parameter, and using least squares minimization between the numerical solution of Eq.~\ref{RayleighOscEqn} and the measured dispersion, we find $f_0=215$ MHz. The fitted resonant frequency is plotted as the blue dashed line in Fig.~\ref{Figure3}, and gives a contact stiffness of $K_2=2.7$~kN/m. We plot the numerical dispersion curve from our fitting as the red solid line in Fig.~\ref{Figure3}. 

Any real solution of Eq.~\ref{RayleighOscEqn} must yield the phase velocity $\omega/k$ smaller than $c_T$, otherwise at least one of the square root terms becomes imaginary. Therefore, the calculated upper dispersion branch terminates at the threshold $\omega=c_T k$, in agreement with the experiment. In some cases leaky wave solutions with complex $\omega$ can be found above the threshold \cite{Garova1999}, however we did not investigate complex solutions since in our experiment the upper branch peak disappeared past the threshold.

Using the oscillator equation of motion, we estimated the relative displacements (and phase) of the microspheres and the surface for various limiting cases \cite{Supplementary}, as is shown in Fig.~\ref{Figure3}. In the flat dispersion region of the lower branch, there is predominantly sphere oscillation with very small surface displacements. Indeed, our experimental data show that the amplitude of the corresponding peak becomes progressively smaller compared to the Rayleigh-like upper branch as the wave vector magnitude is increased~\cite{Supplementary}. 


We estimate the frequency of the contact resonance (and the contact stiffness) of the microspheres using the DMT contact model \cite{DMT1983,Supplementary}, and compare with the resonant frequency obtained from the measured data. The DMT model describes the contact between an elastic sphere and a flat substrate under the presence of adhesive forces. The model assumes that the deformation profile is Hertzian and the adhesive forces act outside the contact area. For the case of small displacements, the full DMT model can be approximated as \cite{DMT1983}:
\begin{equation}
F=KR^{1/2}\alpha^{3/2}-2\pi wR, 
\label{DMT}
\end{equation}
where $F$ is a force applied to the sphere, $\alpha$ is the displacement of the center of the sphere towards the substrate, $w=0.094$ J/$\text{m}^2$ is the work of adhesion between silica and alumina (as the aluminum surface is normally oxidized)~\cite{Supplementary}, $R$ is the effective microsphere radius of contact (which reduces to $R=D/2$ for a sphere on a flat substrate \cite{Hertz}), and $K=(\frac{3}{4}(\frac{1-\nu_s^2}{E_s}+\frac{1-\nu_1^2}{E_1}))^{-1}$ is the the effective modulus, where the aluminum has elastic modulus $E_1=62$ GPa and Poisson's ratio $\nu_1=0.24$~\cite{ASMAluminum}, and the microspheres have elastic modulus $E_s=73$ GPa and Poisson's ratio $\nu_s=0.17$~\cite{GlassHandbook}. Using Eq.~\ref{DMT}, we calculate the equilibrium displacement $\alpha_0=(\frac{2\pi wR^{1/2}}{K})^{2/3}=0.44$~nm, the linearized stiffness around the equilibrium point $K_{2,DMT}=\frac{3}{2}(2\pi w R^2 K^2)^{1/3}=1.1$ kN/m, and $f_{0,DMT}=\frac{1}{2\pi}\sqrt{K_{2,DMT}/m}=140$ MHz. Below the axial contact resonance, a rocking mode is also predicted $f_{rock}=\frac{1}{R^{3/2}}\sqrt{\frac{45w}{4\rho_s}}=10$~MHz~\cite{Peri2005}, however this is significantly below our measured acoustic frequency range. The discrepancy between the estimated and the measured values of $f_{0}$ may be caused by uncertainties in the contact and adhesion models. Challenges in application of DMT to real nanoscale contacts have been underscored by studies in atomic force microscopy~\cite{Bhushan,Hurley2006}. Typical adhesion studies relying on measuring a pull-off force provide limited information for verification of adhesion theories~\cite{Bhushan,Peri2005}. Our experiment provides a direct pathway to the contact stiffness, and thus offers a promising tool for studying adhesion and contact mechanics.  


Potentially useful information can also be obtained by studying the attenuation of the SAWs. As can be seen in Fig.~\ref{Figure2}(d), the peaks in the on-spheres case are broader. Faster attenuation is also evident from a comparison of the signal waveforms in Fig.~\ref{Figure2}(a,b). In the off-spheres case, the acoustic signal decays as the SAW wavepacket leaves the probe spot, with material losses being negligible within the time window of the measurement~\cite{AlexJAP2009}. The apparent decay time is thus determined by the width of the wavepacket (i.e. the width of the excitation spot) and the SAW group velocity. In the on-spheres case the group velocity is lower, yet the decay time is shorter, which indicates that the presence of the spheres leads to additional attenuation. One possible mechanism is scattering due to the spheres. A single oscillator on the surface of a halfspace will radiate acoustic energy into the substrate; however, a collective mode of a periodic array with $\omega<c_Tk$ will not radiate~\cite{Garova1999,AlexJAP2009}. In our case, the sphere packing is not perfectly periodic, which may lead to scattering and radiation into the bulk. In the flat region of the lower dispersion branch, the acoustic mode is close to the contact resonance of the spheres. In this case, the peak width may also be determined by inhomogeneous broadening caused by sphere-to-sphere variation of the contact stiffness. If this is the case, then the peak width can be used to study the statistics of contact stiffnesses. Peak broadening may also be caused by anharmonicity, but we estimate sphere displacements to be in the linear regime~\cite{Supplementary}, and no anharmonic effects such as second harmonic generation are observed. In the future, it would be interesting to study the nonlinear regime, which should be achievable based on observations of a complete detachment of micron-sized particles by high-amplitude SAWs~\cite{MaznevAblation}.  

We have seen that an avoided crossing between the Rayleigh wave and contact resonance of the spheres occurs at wavelengths much larger than the sphere size and is well described by the effective medium approximation. Thus our structure belongs to a class of ``locally resonant metamaterials'', for which interesting effects have been observed on the macroscale~\cite{LiuScience2000,SodaCans}. We expect the effective medium model to break down at shorter wavelengths where phononic crystal effects should be seen~\cite{Garner1993}. Furthermore, our model treats the spheres as independent oscillators that interact through the elastic substrate but not directly. Although the spheres are closed-packed, the model describes the data very well. We believe that this is, again, due to the fact that the acoustic wavelength is large compared to the sphere size. At shorter wavelengths we expect to see rich dynamics due to interaction between the spheres~\cite{NoteOnFytas2012,Tournat2D}; the interaction with SAWs will make such dynamics even richer. 

In summary, we have studied the interaction of SAWs with the contact-based resonance of microspheres forming a two-dimensional granular crystal. The experimental method can be used to study the adhesion and contact mechanics of microparticles. It also enables the study of granular crystals on the microscale. A rich array of dynamical phenomena observed in macroscale granular crystals, and their promise for practical applications~\cite{GranularCrystalReviewChapter}, suggest interesting possibilities for microscale granular crystals. An analogy can also be made between SAWs interacting with a microsphere contact resonance and surface plasmon-polariton waves in a metal film interacting with a localized surface plasmon resonance of a metallic nanoparticle coupled to the metal film~\cite{Smith2012}. This may lead to acoustic analogies of some plasmonic phenomena and applications~\cite{PlasmonicsBook}. Finally, the nonlinearity of the Hertzian contact holds promise for an application of our approach to developing nonlinear SAW devices.   
 
N.B. thanks G. Theocharis for useful discussions, and K. Broderick for guidance in substrate fabrication. This work was supported by the the Defense Threat Reduction Agency through grant HDTRA 1-12-1-0008. A.K. and N.X.F. are also partially supported by NSF grant CMMI-1120724.
%
%


\begin{thebibliography}{99}
\bibitem{NoteSurfaceRipples}
By comparing with measurements made from the front side, we find that the signal contribution from surface ripples is dominant, except in the case of longitudinal waves, where the contribution from refractive index variation is dominant.
\end{thebibliography}
\end{document}


\title{Supplementary Information: Interaction of a Contact Resonance of Microspheres with Surface Acoustic Waves}

\author{N. Boechler$^{1}$, J. K. Eliason$^{2}$, A. Kumar$^{1}$, A. A. Maznev$^{2}$, K. A. Nelson$^{2}$, and N. Fang$^{1}$}

\affiliation{ 
$^1$ Department of Mechanical Engineering, Massachusetts Institute of Technology, Cambridge, MA 02139 \\
$^2$ Department of Chemistry, Massachusetts Institute of Technology, Cambridge, MA 02139 
}

\maketitle

\section{Spectra for Multiple SAW Wavelengths}
\begin{figure}[H]
\begin{center}
\includegraphics[width=17.2cm,height=12.9cm]{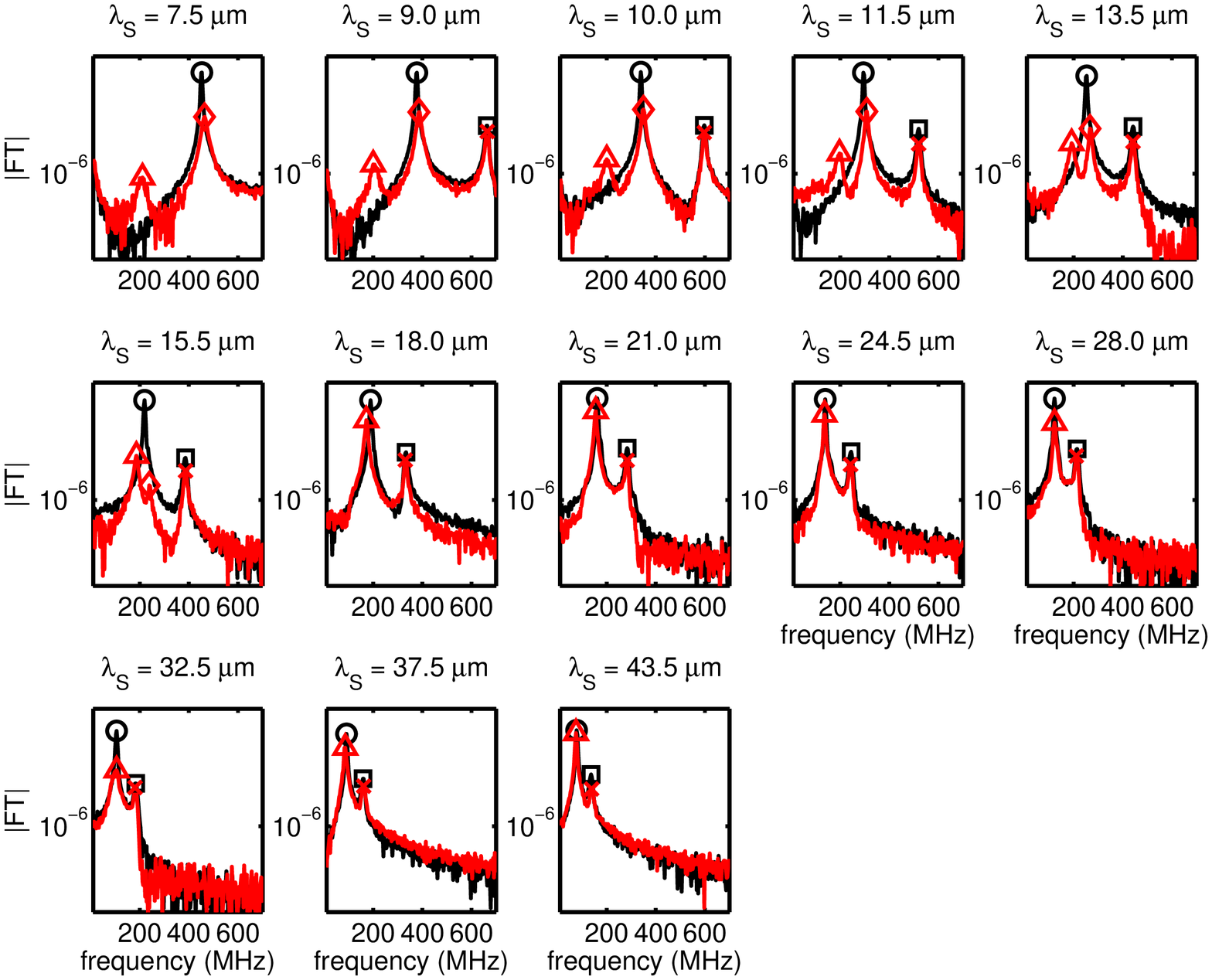}
\end{center}
\caption{\label{SI_Figure3} Measured spectra for multiple SAW wavelengths.} 
\end{figure}

In Fig.~\ref{SI_Figure3}, we show the measured spectra, as in Fig. 2(c) of the main text, for multiple SAW wavelengths. We measure each signal at a $20$ GHz sampling rate, and normalize by the maximum amplitude of the signal. Starting after the sharp initial increase, the slowly decaying component is identified using a smoothing function and subtracted from the normalized signal. Using this signal with the slowly decaying component subtracted, we select a window of $0.24$ $\mu$s duration. This window was then zero-padded to yield a window of $0.36$ $\mu$s total duration, which was used to calculate the FT magnitudes. In some cases, when a small peak is located on the ``shoulder'' of a larger peak, the appearance of the small peak depends of the exact choice of the FT time window: depending on the starting position of the window, the small peak may appear either as a positive peak or a ``dip''. Accordingly, we adjusted the start time (by up to $15$ ns after the sharp initial increase, for each wavelength) to ensure that small peaks appear as positive ones. 

\section{Sample Fabrication Details}
The fused silica wafer was cleaned using Piranha solution prior to aluminum deposition. Following the cleaning, a $0.20$~$\mu$m thick aluminum coating was deposited via electron beam evaporation. The aluminum-coated substrate was then spin-coated with $52$-$14$ photoresist, and diced into $2$~cm square pieces. The photoresist was removed by submerging the substrate in acetone and then isopropanol baths, and dried under N$_2$ flow. 

To assemble the monolayer on the surface, we used a ``wedge-shaped cell'' convective self assembly technique \cite{SunWedgeMethod2010,LopezWedgeMethod2011}. We used $1.08$~$\mu$m diameter silica microspheres, purchased from Corpuscular Inc. as a suspension of $5.0$~wt$\%$ in water. Prior to the deposition, the suspension was further diluted to $1.25$~wt$\%$. Following \cite{SunWedgeMethod2010,LopezWedgeMethod2011}, we submerged the substrate and two glass microscope slides in a $30\%$ hydrogen peroxide bath, at $80^\circ$~C for 1 hour. The substrates and slides were then cleaned with acetone, isopropanol, deionized water, and dried under N$_2$ flow. The substrate was then epoxied on top of one of the glass slides, at a distance from the edge such that, when the second glass slide was clamped on top of this assembly at one edge, the angle of the wedge formed between the substrate and the top glass slide is $3$ degrees. We then deposited $20$~$\mu$L of the diluted suspension on top of the aluminum-coated substrate, and clamped the second glass slide on top (by the edge) to form the wedge. The whole setup was then placed on a tilt, such that the meniscus drying front will receed upwards on an incline of $10$ degrees.

\section{Rayleigh Wave Speed In Layered Substrate}
\begin{figure}[H]
\begin{center}
\includegraphics[width=8.6cm,height=6.45cm]{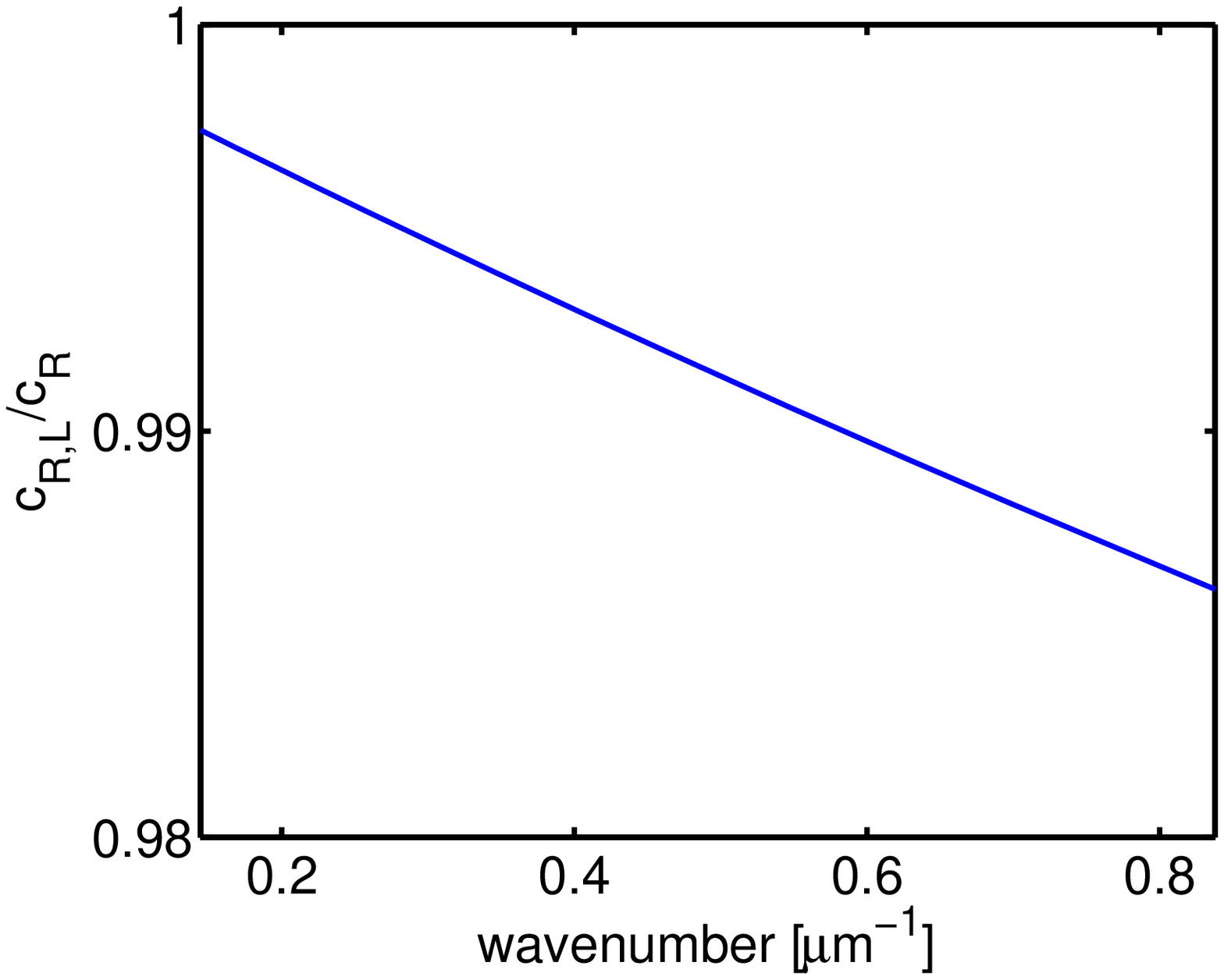}
\end{center}
\caption{\label{SI_Figure4} Rayleigh wave speed $c_{R,L}$ in a fused silica substrate coated with $0.20$~$\mu$m thick aluminum film, normalized by the Rayleigh wave speed in pure fused silica $c_R$, for the experimentally measured wavenumbers.} 
\end{figure}

Following \cite{Ewing}, we calculated the wave speed for a generalized Rayleigh wave in an elastic halfspace (silica) coated with a thin finite layer ($0.20$ $\mu$m thick aluminum) on the top surface. The solution for the layered halfspace allows for higher order surface waves, but for our range of SAW wavelengths we are below the cutoff point for the existence of these additional solutions. Additionally, in this case, the Rayleigh wave speed is dependent on the SAW wavelength. We plot this relationship in Fig.~\ref{SI_Figure4}, for our substrate parameters. At the smallest wavelength (highest difference in wave speed from the pure fused silica halfspace), the wave speed is a maximum of $1.4\%$ slower than in pure fused silica. 

\pagebreak
\section{Derivation of SAW Dispersion Relation for an elastic halfspace with Coupled Surface Oscillators}
Elastic waves in the substrate are described by wave equations $\nabla^2\phi=\frac{1}{c_L^2}\frac{\partial^2\phi}{\partial t^2}$ and $\nabla^2\psi=\frac{1}{c_T^2}\frac{\partial^2\psi}{\partial t^2}$, where $t$ is time, and $\phi$ and $\psi$ are dilational and transverse potentials related to the displacement components $u_x$ and $u_z$ as follows: $u_x=\frac{\partial \phi}{\partial x}-\frac{\partial \psi}{\partial z}$ and $u_z=\frac{\partial \phi}{\partial z}+\frac{\partial \psi}{\partial x}$~\cite{Ewing}. Assuming a traveling wave solution of the form $\phi=f(z)e^{i(\omega t-kx)}$ and $\psi=g(z)e^{i(\omega t-kx)}$, we obtain solutions to the wave equations of the form \cite{Ewing}:  
\begin{equation}
\begin{split}
\phi=B_1e^{kz\sqrt{1-\frac{\omega^2}{k^2c_{L}^2}}+i(\omega t-kx)}
\\
\psi=B_2e^{kz\sqrt{1-\frac{\omega^2}{k^2c_{T}^2}}+i(\omega t-kx)}.
\end{split}
\label{oscsoln}
\end{equation}
The equation of motion for the surface oscillator can be written as: 
\begin{equation}
m\ddot{Z}+K_2(Z-u_{z,0})=0. 
\label{EOM}
\end{equation}
We assume a traveling wave solution for the oscillator motion of the form $Z=e^{i(\omega t-kx)}$, which we substitute Eq.~\ref{EOM} to obtain:
\begin{equation}
Z=\frac{\omega_0^2u_{z,0}}{\omega_0^2-\omega^2}.
\label{OscAmp}
\end{equation}
Substituting Eq.~\ref{OscAmp} into the boundary conditions at $z=0$ (shown in Eq. 1 of the main text), we obtain:
\begin{equation}
\begin{split}
\sigma_{zz}=\frac{K_2}{A}(\frac{\omega^2u_{z,0}}{\omega_0^2-\omega^2}) 
\qquad
\sigma_{xz}=0. 
\end{split}
\label{newBCs}
\end{equation}
Using Eq.~\ref{oscsoln}, Eq.~\ref{newBCs}, and isotropic linear elastic stress-strain relations \cite{Ewing}, we obtain the dispersion relation for the SAWs in our coupled oscillator system, as shown in Eq. 2 of the main text. Additionally, we note that Eq. \ref{OscAmp} was also used to estimate the relative oscillator and surface displacements (and their relative phase), as shown in Fig. 3 of the main text.

\section{Work of Adhesion at a Silica-Alumina Interface}
We estimate the work of adhesion $w$ between our silica microspheres and the aluminum-coated silica substrate. As the aluminum-coated substrate has a thin oxidized layer (alumina) on top of the aluminum, we calculate the work of adhesion at a silica-alumina interface. Following \cite{Israelachvili}, we find the work of adhesion in terms of the Hamaker constant $A_{12}$ for a silica-alumina interface, where $w=\frac{A_{12}}{12\pi D_0^2}$ ($D_0=0.165$~nm is a standard value used for the interfacial cutoff separation distance for a variety of media \cite{Israelachvili}). This coefficient takes into account the van der Waals forces between the two surfaces in contact. The Hamaker constant can be defined in terms of the material properties of the two materials in contact, interacting across a third medium (in this case air), such that:
\begin{equation}
A_{12}=\frac{3kT}{4}(\frac{\epsilon_1-\epsilon_3}{\epsilon_1+\epsilon_3})(\frac{\epsilon_2-\epsilon_3}{\epsilon_2+\epsilon_3})+\frac{3h\nu_e}{8\sqrt{2}}\frac{(n_1^2-n_3^2)(n_2^2-n_3^2)}{(n_1^2+n_3^2)^{1/2}(n_2^2+n_3^2)^{1/2}[(n_1^2+n_3^2)^{1/2}+(n_2^2+n_3^2)^{1/2}]}, 
\label{Hamaker}
\end{equation}
where subscripts $1$, $2$, and $3$ correspond to silica, alumina, and air, respectively, $T=293$ K is room temperature, $\nu_e=3.2\cdot10^{15}$ Hz is the main electronic absorption frequency in the UV, $k$ is Boltzmann's constant, $h$ is Planck's constant, permittivities $\epsilon_1=3.8$~F/m, $\epsilon_2=11$~F/m, and $\epsilon_3=1.0$~F/m, and refractive indices $n_1=1.45$, $n_2=1.75$, and $n_3=1.00$. All of the preceding constants are tabulated in \cite{Israelachvili}. Using Eq.~\ref{Hamaker}, we calculate $A_{12}=9.6\cdot10^{-20}$ J. This gives a work of adhesion of $w=0.094$ J/$\text{m}^2$.

\section{Contact Model}
There are several models that are commonly used to describe the contact of elastic spheres under the presence of adhesive forces \cite{Bhushan}. This includes the DMT \cite{DMT1983}, Jordan-Kendall-Roberts (JKR) \cite{JKR1971}, and Maugis-Dugdale \cite{Maugis1992} models. In the main text, we used the DMT model to estimate the frequency of the contact resonance, which assumes that the deformation profile is Hertzian and the adhesive forces act outside the contact area \cite{DMT1983}. In the JKR model, the deformation profile is not constrained to be Hertzian, and the adhesive forces are assumed to act only within the contact area \cite{JKR1971}. The DMT model is typically applied to weakly adhesive systems with small, stiff particles, and the JKR model to strongly adhesive systems with large, soft particles \cite{Bhushan}. The more general Maugis-Dudgale model continously transitions between these two regimes, and can be written as follows \cite{Bhushan,Maugis1992}:
\begin{equation}
\begin{split}
1=\frac{\lambda a^2}{2}(\frac{K}{\pi R^2 w})^{2/3}[\sqrt{M^2-1}+(M^2-2)(\text{arctan}\sqrt{M^2-1})]\\
+\frac{4\lambda^2 a}{3}(\frac{K}{\pi R^2 w})^{1/3}[1-M+\sqrt{M^2-1}(\text{arctan}\sqrt{M^2-1})],\\
\alpha=\frac{a^2}{R}-\frac{4\lambda a}{3}(\frac{\pi w}{RK})^{1/3}\sqrt{M^2-1},\\
F=\frac{Ka^3}{R}-\lambda a^2(\frac{\pi wK^2}{R})^{1/3}[\sqrt{M^2-1}+M^2(\text{arctan}\sqrt{M^2-1})],
\label{Maugis}
\end{split}
\end{equation}
where $a$ is the radius of contact, $M$ is the width of an annular region where adhesive forces are assumed to act, $z_0=0.4$~nm is the approximate interatomic distance at the interface \cite{Israelachvili}, and $\lambda=\frac{2.06}{z_0}(\frac{Rw^2}{\pi K^2})^{1/3}$ is a non-dimensional parameter that describes the behavior of the contact. Equation~\ref{Maugis} approaches the JKR model as $\lambda \rightarrow \infty$, and the DMT model as $\lambda \rightarrow 0$. For our silica microspheres in contact with an aluminum-coated substrate, we calculate $\lambda=0.5$. We estimate the frequency of the contact resonance using Eq.~\ref{Maugis}, and find it to be less than $5\%$ smaller than the frequency predicted using the DMT contact model.

\section{Surface Displacement Amplitude}
Following \cite{Maznev1990}, we estimate the displacement amplitude in the laser-generated SAW to be $u_{z,0}=g\Gamma Q_0=23$~pm, where:
\begin{equation}
g=\frac{3\alpha_L(1-R_{ref})}{\rho_1Cp}(1-\frac{4c_T^2}{3c_L^2}),
\label{DispThermal2}
\end{equation}
\begin{equation}
\Gamma=[(c_T^2-\frac{1}{2}c_R^2)[(c_T^2-c_R^2)^{-1}+(c_L^2-c_R^2)^{-1}]-2]^{-1},
\label{DispThermal3}
\end{equation}
$Q_0=Q_L/(\pi R_L^2)$ is the laser energy density (with a pulse energy of $Q_L=2.44$~$\mu$J and spot radius of $R_L=150$~$\mu$m), and $\alpha_L=23.6$~$\mu$m/m$\cdot$K is the linear thermal expansion coefficient, $\rho_1=2.7$ g/cm$^3$ is the density, $C_p=0.90$~kJ/kg$\cdot$K is the specific heat, and $R_{ref}=0.92$ is the surface reflectivity of aluminum~\cite{ASMAluminum}. Although the microsphere vibration amplitude can be greater than the substrate surface displacement, this value is over an order of magnitude less than the equilibrium displacement of the microsphere due to adhesive forces, and thus serves as a basis for assuming the linearity of the response.